\documentclass[12pt,twoside]{article}\textheight 620pt\textwidth 6.5in
 \usepackage{amssymb} 
\begin{document}

 \newtheorem{conjecture}{Conjecture} \newtheorem{lemma}{Lemma}
 \newtheorem{notation}{Notation} \newtheorem{theorem}{Theorem}
 \newtheorem{definition}{Definition} \newtheorem{remark}{Remark}
 \newtheorem{proposition}{Proposition}

 \title{The Tale of One-Way Functions\thanks
 {Problemy Peredachi Informatsii 39(1), 2003 \
 ( = Problems of Information Transmission),\hfil\mbox{}
 UDK~621.391:519.2}}

 \author{Leonid A. Levin\thanks
        {This research was partially conducted by the author
 for the Clay Mathematics Institute and supported by the Institut
 des Hautes \'Etudes Scientifiques and NSF, grant no.~CCR-9820934.}\\
 Boston University\thanks
 {Computer {Sci.} dept., 111 Cummington St., Boston, MA 02215.}}

 \date{} \maketitle \begin{flushright}\begin{tabular}{l}
 \emph {All the king's horses, and all the king's men,}\\*
 \emph {Couldn't put Humpty together again.}
 \end{tabular}\end{flushright}

\begin{abstract} The existence of one-way functions (owf) is arguably
the most important problem in computer theory. The article discusses and
refines a number of concepts relevant to this problem. For instance, it
gives the first combinatorial complete owf, i.e., a function which is
one-way if any function is. There are surprisingly many subtleties in
basic definitions. Some of these subtleties are discussed or hinted at
in the literature and some are overlooked. Here, a~unified approach is
attempted. \end{abstract}

\section{Introduction I: Inverting Functions}

From time immemorial, humanity has gotten frequent, often cruel,
reminders that many things are easier to do than to reverse. When the
foundations of mathematics started to be seriously analyzed, this
experience immediately found a formal expression.

\subsection{An Odd Axiom}

Over a century ago, George Cantor reduced all the great variety of
mathematical concepts to just one---the concept of sets---and derived
all mathematical theorems from just one axiom scheme---\emph{Cantor's
Postulate}. For each set-theoretical formula $A(x)$, it postulates the
existence of a set containing those and only those $x$ satisfying $A$.
This axiom looked a triviality, almost a definition, but was soon found
to yield more than Cantor wanted, including contradictions. To salvage
its great promise, Zermelo, Fraenkel, and others pragmatically replaced
Cantor's Postulate with a collection of its restricted cases, limiting
the types of allowed properties $A$. The restrictions turned out to
cause little inconvenience and precluded (so far) any contradictions;
the axioms took their firm place in the foundations of mathematics.

In 1904, Zermelo noticed that one more axiom was needed to derive all
known mathematics, the (in)famous Axiom of Choice: every function $f$
has an inverse $g$ such that $f(g(x))=x$ for $x$ in the range of $f$. It
was accepted reluctantly; to this day, proofs dependent on it are being
singled out. Its strangeness was not limited to going beyond Cantor's
Postulate---it brought paradoxes! Allow me a simple illustration based
on the ability of the axiom of choice to enable a \emph{symmetric}
choice of an arbitrary integer.

Consider the additive group $\mathbb{T}=\mathbb{R}/\mathbb{Z}$ of reals
mod $1$ as points $x\in[0,1)$ on a circle; take its subgroup
$Q_{10}\subset\mathbb{T}$ of decimal fractions $a/10^b$. Let $f(x)$ be
the (countable) coset $x+Q_{10}$, i.e., $f$ projects $\mathbb{T}$ onto
its factor group $\mathbb{T}/Q_{10}$. Any inverse $g$ of $f$ then
selects one representative from each coset. Denote by
$G=g(f(\mathbb{T}))$ the image of such a $g$; then each $x\in\mathbb{T}$
is brought into $G$ by exactly one rational shift $x+q$, $q\in Q_{10}$.
Now I will deviate from the standard path to emphasize the elementary
nature of the paradox. One last notation: $q'= (10 q)\bmod 1$ is $q\in
Q_{10}$, shortened by the removal of its most significant digit.

For a pair $p,q\in Q_{10}$, I bet $2:1$ that $x+p$ rather than $x+q$
falls in $G$ for a random $x\in\mathbb{T}$. The deal should be
attractive to you since my bet is higher while conditions to win are
completely symmetric under rotation of $x$. To compound my charitable
nature to its extreme, I offer such bets for all $q\in Q_{10}$, $q>0$,
$p=q'$, not just one pair. If you rush to accept, we choose a random $x$
(by~rolling dice for all of its digits) and find the unique ${\bf q}\in
Q_{10}$ for which $x+{\bf q}\in G$. Then I lose one bet for this ${\bf
q}$ and win ten (for each $q$ such that $q'={\bf q}$). Generosity pays!

This paradox is not as easy to dismiss as is often thought. Only 11 bets
are paid in each game: no infinite pyramids. Moreover, if $x$ is drawn
from a sphere $S_2$, a finite number of even unpaid bets suffices:
Banach and Tarski \cite{banach} constructed 6 pairs, each including a
set $A_i\subset S_2$ and a rotation~$T_i$; betting $x\in A_i$ versus
$T_i(x)\in A_i$, they lose one bet and win two for each $x$. Our $x+p$
and $x+q$ above are tested for the same condition and differ in finitely
many digits; all digits are evenly distributed and independent. One can
refuse the thought experiment of rolling the infinite number of digits
of $x$ or the question of whether $x+q\in G$, but this amounts to
rejecting basic concepts of the set theory. It is simpler to interpret
the refusal to bet as a hidden disbelief in the Axiom of Choice.

\subsection{Finite Objects: Exhaustive Search}

These problems with inverting functions have limited relevance to
computations. The latter deal with finite objects, which are naturally
well-ordered by Induction Axioms, rendering the Axiom of Choice
unnecessary. There are other mitigating considerations, too. Shannon
Information theory assigns to a random variable $x$ as much information
about its function $f(x)$ as to $f(x)$ about $x$. Kolmogorov
(algorithmic) information theory (see~\cite{Kolm1,ku87}) extended this
concept from random to arbitrary $x$: $I(x:y)$ is the difference between
the smallest lengths of programs generating $y$ and of those
transforming $x$ to $y$. Kolmogorov and I proved in 1967 that this
measure is symmetric too, like Shannon's, albeit approximately
\cite{zl}.

The proof involved a caveat: computationally prohibitive exhaustive
search of all strings of a given length. For instance, the product $pq$
of two primes contains as much (i.e., all) information about them as
vice versa, but \cite{RSA} and great many other things in modern
cryptography depend on the assumption that recovering the factors of
$pq$ is infeasible. Kolmogorov suggested at the time~\cite{K67} that
this information symmetry theorem may be a good test case to prove that
for some tasks exhaustive search cannot be avoided (in today's terms,
${\rm P}\ne{\rm NP}$).

The RSA application marked a dramatic twist in the role of the inversion
problem: from notorious troublemaker to priceless tool. RSA was the
first of the myriad bewildering applications which soon followed. At the
heart of many of them was the discovery that the hardness of every
one-way function $f(x)$ can be focused into a \emph{hard-core} bit,
i.e., an easily computable predicate $b(x)$ which is as hard to
determine from $f(x)$, or even to guess with any noticeable correlation,
as to recover $x$ completely.

In \cite{BM}, the first hard-core for $f(x)= a^x\bmod p$ was found,
which was soon followed with hard-cores for RSA and Rabin's function
$(x^2\bmod n)$, $n=pq$, and for ``grinding'' functions
$f^*(x_1,\ldots,x_n)= f(x_1),\ldots, f(x_n)$ (see \cite{yao}; for the
proof of Isolation Lemma used in \cite{yao}, see \cite{ow}). In
\cite{GL}, the general case was proved (see also \cite {L93,Knuth}).

The importance of such hard-cores comes from their use, proposed in
\cite{BM,yao} for unlimited deterministic generation of perfectly random
bits from a small random seed $s$. In the case of permutation $f$, such
generators are straightforward: $g_s(i)=b(f^i(s))$,
$i=0,1,2,\ldots\strut$. The general case was worked out in \cite{HILL}.

With the barrier between random and deterministic processes thus broken,
many previously unthinkable feats were demonstrated in the 80s. Generic
cryptographic results (such as, e.g., \cite{NY}), zero-knowledge proofs,
and implementation of arbitrary protocols between distrusting parties
(e.g., card games) as full information games \cite{GMW} are just some of
the many famous examples. This period was truly a golden age of Computer
Theory brought about by the discovery of the use of one-way functions.

\section{Introduction II: Extravagant Models}

\subsection{The Downfall of RSA}

This development was all the more remarkable as the very existence of
one-way (i.e., easy to compute, infeasible to invert) functions remains
unproven and subject to repeated assaults. The first came from Shamir
himself, one of the inventors of the RSA system. He proved \cite{shamir}
that factoring (on infeasibility of which RSA depends) can be done in
polynomial number of arithmetic operations. This result uses a so-called
``unit-cost model,'' which charges one unit for each arithmetic
operation, however long the operands. Squaring a number doubles its
length, repeated squaring brings it quickly to cosmological sizes.
Embedding a huge array of ordinary numbers into such a long one allows
one arithmetic step to do much work, e.g., to check exponentially many
factor candidates. The closed-minded cryptographers, however, were not
convinced and this result brought a dismissal of the unit-cost model,
not RSA.

Another, not dissimilar, attack is raging this very moment. It started
with the brilliant result of Peter Shor. He factors integers in
polynomial time using an imaginary analog device, Quantum Computer (QC),
inspired by the laws of quantum physics taken to their formal extreme.

\subsection{Quantum Computers}

QC has $n$ interacting elements, called q-bits. A pure state of each is
a unit vector on the complex plane $\mathbb{C}^2$. Its two components
are quantum amplitudes of its two Boolean values. A state of the entire
machine is a vector in the tensor product of $n$ planes. Its $2^n$
coordinate vectors are tensor-products of q-bit basis states, one for
each $n$-bit combination. The machine is cooled, isolated from the
environment nearly perfectly, and initialized in one of its basis states
representing the input and empty memory bits. The computation is
arranged as a sequence of perfectly reversible interactions of the
q-bits, putting their combination in the superposition of a rapidly
increasing number of basis states, each having an exponentially small
amplitude. The environment may intervene with errors; the computation is
done in an error-correcting way, immune to such errors as long as they
are few and of special restricted forms. Otherwise, the equations of
Quantum Mechanics are obeyed with unlimited precision. This is crucial
since the amplitudes are exponentially small and deviations in remote
(hundredth or even much further) decimal places would overwhelm the
content completely. In \cite{shor}, such computers are shown capable of
factoring in polynomial time. The exponentially many coordinates of
their states can, using a rough analogy, explore one potential solution
each and concentrate the amplitudes in the one that works.

\subsection{Small Difficulties}

There are many problems with such QCs. For instance, thermal isolation
cannot be perfect. Tiny backgrounds of neutrinos, gravitational waves,
and other exotics, cannot be shielded. Their effects on quantum
amplitudes need not satisfy the restrictions on which error-correcting
tools depend. Moreover, nondissipating computing gates, even classical,
remain a speculation. Decades past, their existence was cheerfully
proclaimed and even proved for worlds where the laws of physical
interaction can be custom-designed. In our world, where the
electromagnetic interaction between electrons, nuclei, and photons is
about the only one readily available, circuits producing less entropy
than computing remain hypothetical. So, low temperatures have limits and
even a tiny amount of heat can cause severe decoherence problems.
Furthermore, the uncontrollable degrees of freedom need not behave
simply as heat. Interaction with the intricately correlated q-bits may
put them in devilish states capable of conspiracies which defy
imagination.

\subsection{Remote Decimals}

All such problems, however, are peanuts. The major problem is the
requirement that basic quantum equations hold to multi-hundredth if not
millionth decimal positions where the significant digits of the relevant
quantum amplitudes reside. We have never seen a physical law valid to
over a dozen decimals. Typically, every few new decimal places require
major rethinking of most basic concepts. Are quantum amplitudes still
complex numbers to \emph{such} accuracies or do they become quaternions,
colored graphs, or sick-humored gremlins? I suspect physicists would
doubt even the laws of arithmetic pushed that far. In fact, we know that
the most basic laws cannot all be correct to hundreds of decimals: this
is where they stop being consistent with each other!

And what is the physical meaning of 500 digit long numbers? What could
one possibly mean by saying ``This box has a remarkable property: its
many q-bits contain the Ten Commandments with the amplitude whose first
500 decimal places end with 666''? What physical interpretation could
this statement have even for just this one amplitude? Close to the
tensor product basis, one might have opportunities to restate the
assertions using several short measurable numbers instead of one long.
Such opportunities may also exist for large systems, such as lasers or
condensates, where individual states matter little. But QC factoring
uses amplitudes of an exponential number of highly individualized basis
states. I doubt anything short of the most generic and direct use of
these huge precisions would be easy to substitute. One can make the
amplitudes more ``physical'' by choosing a less physical basis. Let us
look into this.

\subsection{Too Small Universe}\label{small-u}

QC proponents often say they win either way, by making a working QC or
by finding a correction to Quantum Mechanics; e.g., in \cite{shor-spr}
Peter Shor said: ``If there are nonlinearities in quantum mechanics
which are detectable by watching quantum computers fail, physicists will
be VERY interested (I would expect a Nobel prize for conclusive evidence
of this).''

Consider, however, this scenario. With few q-bits, QC is eventually made
to work. The progress stops, though, long before QC factoring starts
competing with pencils. The QC people then demand some noble prize for
the correction to the Quantum Mechanics. But the committee wants more
specifics than simply a nonworking machine, so something like observing
the state of the QC is needed. Then they find the Universe too small for
observing individual states of the needed dimensions and accuracy.
(Raising sufficient funds to compete with pencil factoring may justify a
Nobel Prize in Economics.)

Let us make some calculations. In cryptography, the length $n$ of the
integers to factor may be a thousand bits (and could easily be
millions.) By $\sim n$, I will mean a reasonable power of $n$.
A~$2^{\sim n}$~dimensional space $H$ has $2^{2^{\sim n}}$ nearly
orthogonal vectors. Take a generic $v\in H$. The minimal size of a
machine which can recognize or generate $v$ (approximately) is
$K=2^{\sim n}$---far larger than our Universe. This comes from a
cardinality argument: $2^{\sim K}$ machines of $K$ atoms. Let us call
such $v$ ``megastates.''

There is a big difference between untested and untestable regimes.
Claims about individual megastates are untestable. I can imagine a
feasible way to separate any \emph{two} QC states \emph{from each
other}. However, as this calculation shows, no machine can separate a
generic QC state from the set of all states more distant from it than QC
tolerates. So, what thought experiments can probe the QC to be in the
state described with the accuracy needed? I would allow to use the
resources of the entire Universe, but \emph{not more}!

Archimedes made a great discovery that digital representation of numbers
is exponentially more efficient than analog ones (sand pile sizes). Many
subsequent analog devices yielded unimpressive results. It is not clear
why QCs should be an exception.

\subsection{Metric versus Topology}\label{metr}

A gap in quantum formalism may be contributing to the confusion.
Approximation has two subtly different aspects: metric and topology.
Metric tells how close our ideal point is to a specific wrong one.
Topology tells how close it is to the combination of all unacceptable
(nonneighboring) points. This may differ from the distance to the
closest unacceptable point, especially for quantum systems.

In infinite dimensions, the distinction between 0 and positive
separation between a point and a set varies with topologies. In finite
dimensions, 0-vs.-positive distinction is too coarse: all topologies
agree. Since $2^{500}$ is finite only in a very philosophical sense, one
needs a quantitative refinement, some sort of a weak-topological (not
metric) \emph{depth} of a neighborhood polynomially related to resources
required for precision to a given depth. Then, precision to reasonable
depths would be physical, e.g., allow one to generate points inside the
neighborhood, distinguish its center from the outside, etc.

Metric defines $\varepsilon$-neighborhoods and is richer in that than
topology where the specific value of $\varepsilon$ is lost (only
$\varepsilon>0$ is assured). However, metric is restricted by the axiom
that the intersection of any set of $\varepsilon$-neighborhoods is
always another $\varepsilon$-neighborhood. Quantum proximity may require
both benefits: defined depth $\varepsilon$ and freedom to express it by
formulas violating the ``intersection axiom.'' Here is an example of
such a violation, without pretense of relevance to our needs. Suppose a
neighborhood of 0 is given by a set of linear inequalities $f_i(x)<1$;
then its depth may be taken as $1/\sum\limits_i\|f_i\|$. Restricting $x$
to the unit sphere would render this depth quadratically close to metric
depth. A more relevant formula may need preferred treatment of tensor
product basis.

\subsection{The Cheaper Boon}

QC of the sort that factors long numbers seems firmly rooted in science
fiction. It is a pity that popular accounts do not distinguish it from
much more believable ideas, like Quantum Cryptography, Quantum
Communications, and the sort of Quantum Computing that deals primarily
with locality restrictions, such as fast search of long arrays. It is
worth noting that the reasons why QC must fail are by no means clear;
they merit thorough investigation. The answer may bring much greater
benefits to the understanding of basic physical concepts than any
factoring device could ever promise. The present attitude is analogous
to, say, Maxwell selling the Daemon of his famous thought experiment as
a path to cheaper electricity from heat. If he did, much of insights of
today's thermodynamics might be lost or delayed.

The rest of the article ignores any extravagant models and stands fully
committed to the Polynomial Overhead Church--Turing Thesis: Any
computation that takes $t$ steps on an $s$-bit device can be simulated
by a Turing Machine in $s^{O(1)}t$ steps within $s^{O(1)}$ cells.

\section{The Treacherous Averaging}

Worst-case hardness of inverting functions may bring no significant
implications. Imagine that all instances come in two types: ``easy'' and
``hard.'' The easy instances $x$ take $\|x\|^2$ time. An~exponential
expected time of randomized algorithms is required \emph{both} to solve,
and \emph{to find} any hard instance. So, the Universe would be too
small to ever produce intractable instances, and the inversion problem
would pose no practical difficulty. It is ``generic,'' not worst-case,
instances that both frustrate algorithm designers and empower
cryptographers to do their incredible feats. The definition of
``generic,'' however, requires great care.

\subsection{Las Vegas Algorithms}\label{las}

First we must agree on how to measure the performance of inverters.
Besides instances $x=f(w)$, algorithms $A(x,\alpha)$ inverting one-way
functions $f$ can use random dice sequences
$\alpha\in\{0,1\}^{\mathbb{N}}$. They never need a chance for (always
filterable) wrong answers. So we restrict ourselves to \emph{Las Vegas}
algorithms which can only produce a correct output, abort, or diverge.

For a given $x$, the performance of $A$ has two aspects: the
volume\footnote {I say \emph{volume} rather than \emph{time}, for
greater robustness in the case of massively parallel models.}
 $V$ of computation and the chance $p_V$ (over $\alpha$) of success in
$V$ steps. They are not independent: $p$ can be always boosted while
roughly preserving $V/p$ (more precisely $-V/\log(1\!-\!p)$)
 by simply running $A$ on several independent $\alpha$. This idea
suggests the popular requirement that Las Vegas algorithms be normalized
to, say, $p\ge \frac12$. The problem with this restriction is that
estimating $p$ and the needed number of trials may require exponential
volume overhead in the worst case. Thus, only such measures as average
volume can be kept reasonable while normalizing the chance. It is
important that both are averaged only over $A$'s own dice $\alpha$; the
instance $x$ is chosen by the adversary. In this setting, the duplicity
of performance aspects does become redundant: $p$ and average volume are
freely interchangeable (to shrink the latter, one simply runs $A$ with a
small chance).

Combining several runs into one lessens the modularity and counting the
runs needed does involve some overhead. However, there are more
substantial reasons to prefer normalization of average volume to that of
the success rate $p$. In some settings, success is a matter of degree.
For instance, different inverses of the same instance of a owf may be of
different and hard-to-compare value. Normalizing the average volume, on
the other hand, is robust. This volume bound may be $O(1)$ if the model
of computation is very specific. If flexibility between several
reasonable models is desired, polynomial bounds, specific to each
algorithm, may be preferable. There is one obstacle: the set of
algorithms with a restricted expected complexity is not recursively
enumerable. We can circumvent this problem by using the following
enforceable form of the bound.

\begin{definition}
\emph{Las Vegas} algorithms $A(x,\alpha)\in\mathop{\rm LV}(b)$ start
with a given bound $b(x)$ on expected computation volume. At any time,
the algorithm can bet a part of the remaining volume, so that it is
doubled or subtracted depending on the next dice of $\alpha$.
$\mathop{\rm LV}(O(1))$ usually suffices and we will abbreviate it as
L.\footnote{Pronounced ``Las'' algorithms to hint at Las Vegas and the
term's inventor Laszlo Babai. I would like to stress that no YACC (Yet
Another Complexity Class) is being introduced here. L is a \emph{form}
of algorithms; this is much less abstract than a class of algorithms or,
especially, a class of problems solvable by a class of algorithms.
Besides, it is not really new, just a slight tightening of the Las Vegas
restriction.} \end{definition}

Despite its tight $O(1)$ expected complexity bound, L is robust since
any Las Vegas algorithm can be put in this form, roughly preserving the
ratio between the complexity bound and the success rate. The inverse of
the latter gives the number of runs required for a constant chance of
success, thus playing the role of running time. An extra benefit is that
a reader adverse to bothering with the inner workings of computers can
just accept their restriction to L and view all further analysis in
purely probabilistic terms!

\subsection{Multimedian Time}\label{mult}

Averaging over the instance $x$ is, however, much trickier. It is not
robust to define generic complexity of an algorithm $A(x)$ running in
$t(x)$ steps as its expected time ${\bf E}_x t(x)$. A different device
may have a quadratic time overhead. For instance, recognition of input
palindromes requires quadratic time on a Turing machine with one tape,
but only linear time with two tapes. It may be that a similar overhead
exists for much slower algorithms, too. Then $t(x)$ may be, say,
$\|x\|^2$ for $x\not\in0^*$, while $t(0\ldots0)$ may be $2^{\|x\|}$ for
one device and $4^{\|x\|}$ for another. Take $x$ uniformly distributed
on $\{0,1\}^n$. Then ${\bf E}_x t(x)$ for these devices would be
quadratic and exponential respectively: averaging does not commute with
squaring. Besides, this exponential average hardness is misleading since
the hard instances would never appear in practice!

More device-independent would be the \emph{median} time, the minimal
number of steps spent for the \emph{harder half} of instances. This
measure, however, is not robust in another respect: it can change
dramatically if its \emph{half} threshold is replaced with, say, a
\emph{quarter}.

Fortunately, these problems disappear as one of the many benefits of our
Las Vegas conventions. One can simply take algorithms in L and measure
their chance of success for inputs chosen randomly with a given
distribution. The inverse of this chance, as a function of, say, input
length, is a robust measure of \emph{security} of a one-way function.
This measure is important in cryptography, where any noticeable chance
of breaking the code must be excluded. A different measure is required
for positive tasks aimed at success for almost all instances. We start
by considering a combined distribution over all instance lengths.

\begin{definition}
We consider an L-distribution of instances, i.e., a distribution of
outputs of an L-algorithm's on empty input.\footnote{If the instance
generator is not algorithmic, the definition can be modified to use the
output length instead of complexity in the definition of L. The
instances of length $n$ should have probabilities combining to a
polynomial, e.g., $p(x)=1/(\|x\|\log\|x\|)^2$.} Now, we run the
generator $k$ times (spending an average time of $O(k)$) and apply the
inverter until all generated solvable instances are solved.\footnote{If
the generator can produce unsolvable instances too, the definition
ignores them.} The number of trials is a random variable depending on
the inverter's and generator's dice. Its median value MT$(k)$ is called
the \emph{multimedian} time of inverting $f$ by $A$. \end{definition}

This measure is robust in many respects. It commutes with squaring the
inverter's complexity and, thus, is robust against variation of models.
It does not depend much on the $\frac12$ probability cut-off used for
median. Indeed, increasing $k$ by a factor of $c$ raises MT$(k)$ as much
as does tightening the inverter's failure probability to $2^{-c}$.

MT is relevant for both upper and lower bounds. Let T$(x)$ be high for
$\varepsilon$ fraction of $x\in\{0,1\}^n$. Then MT$(k)$ is as high for
$k=n^3/\varepsilon$. Conversely, let MT$(k)$ be high. Then, with
overwhelming probability, T$(x_i)$ is at least as high for some of
$n=k^2$ random $x_1,\ldots, x_n$ (and $\sum\limits_i\|x_i\|=O(n)$).

\subsection{Nice Distributions}\label{round}

So far, we addressed the variance of performance of a randomized
algorithm over its variable dice for a fixed input, as well as the issue
of averaging it over variable input with a given distribution. Now we
must address the variance of distributions. Choosing the right
distributions is not always trivial, a fact often dismissed by declaring
them uniform. Such declarations are confusing, though, since many
different distributions deserve the name.

For instance, consider graphs $G=(V,E\!\subset\! V^2)$, $\|V\|=n$, where
$n$ is chosen with probability $c/n^2$ (here, $c$ is a normalizing
factor). For a given $n$, graphs $G$ are chosen with two distributions,
both with a claim to uniformity: $\mu_1$ chooses $G$ with equal
probability among all $2^{n^2}$ graphs; $\mu_2$ first chooses $k=\|E\|$
with uniform probability $1/(n^2+1)$ and then $G$ with equal probability
among all the $C^k_{n^2}$ candidates. The set $\{G:\: \|E\|=n^{1.5}\}$
has then $\mu_2$ probability $1/(n^2+1)$, while its $\mu_1$ is
exponentially small. In fact, all nice distributions can be described as
uniform in a reasonable representation. Let me reproduce the argument,
sketched briefly in \cite{rp}, adding an additional aspect that I will
use later.

Let us use set-theoretic representation of integers: $n=\{0,1,\ldots,
n-1\}$. A measure $\mu$ is an additive real function of \emph{sets} of
integers; $\mu(n)=\mu(\{0\})+\mu(\{1\})+\ldots+\mu(\{n-1\})$ is its
monotone \emph{distribution} function. Its \emph{density} $\mu'(n)=
\mu(n+1)-\mu(n)=\mu(\{n\})$ is the probability of $\{n\}$ as a
singleton, rather than of a set $n=\{0,1,\ldots,n-\!1\}$. Let $Q_2$ be
the set of binary fractions $i/2^{\|i\|}\in[1/2,1)$. We round the
real-valued $\mu$ to $Q_2$, keeping only as many binary digits as needed
for constant factor accuracy of probabilities.

\begin{definition}
$\mu\colon\mathbb{N}\to Q_2$ is \emph{perfectly rounded} if $\mu(x)$ is
the shortest fraction within the interval
$\bigl(\mu(x-1),\mu(x+1)\bigr)$ and $-\log\mu(\{x\})=O(\|x\|)$.
\end{definition}

The last condition is just a convenience and can be met simply by mixing
the (monotone) $\mu$ with some simple distribution.

\begin{lemma}\label{rnd}
Every computable $\mu\colon\mathbb{N}\to Q_2$ can be uniformly
transformed into a perfectly rounded $\mu_1$ computable at most $\|x\|$
times slower than $\mu(x)$ and so that, for nondecreasing $\mu$ (i.e.,
one with $\mu'\ge 0$), $\mu'_1\ge\mu'/4$. \end{lemma}

Monotonicity is assured by comparing $\mu(x)$ with $\mu(y)$, for
prefixes $y$ of $x\in Q_2$. Then the claim can be achieved by rounding.
First, round $\mu(x)$ to the shortest binary $p$ that is closer to it
than to any other $\mu(y)$ and call these rounded values \emph{points}.
Find all \emph{slots}, i.e., closest to $p$ shorter binary fractions of
each binary length. Then, for each slot in the order of increased
lengths, find the point that fills it in the successive roundings until
the slot for $x$ is found.

All perfectly rounded $\mu$ have a curious property: both $m(x)=
\mu(x)/\mu(\{x\})$ and $\|m(x)\|=-\log\mu(\{x\})$ are always integers,
making $\mu(x)={.}m(x)\in Q_2$, a fraction expressed by bits of $m(x)$
following the dot. So $m$ is quite uniformly distributed:
$2k\mu(m^{-1}(k))\in[1,2]$ for $k\in m(\mathbb{N})$. It~is also
computable in polynomial time, as $m^{-1}$ is (by binary search). So, we
can use $m(x)$ as an alternative representation for $x$, in which the
distribution $\mu$ is remarkably uniform.

Simple distributions are not normally general enough. They may be the
ultimate source of the information in the instances $x$ of our problems,
but the original information $r$ is transformed into~$x$ by some process
$A$ that may itself be something like a one-way function. We can assume
that~$A$ is an algorithm with a reasonable time bound, but not that its
output distribution is simple. Such distributions are called
\emph{samplable}. In \cite{samp}, samplable distributions are dealt with
in a similar manner as with those in this section, though through a
different trick.

\section{Completeness}

\subsection{Complete Distributions and Inverters}

Given a function to invert, how would one generate hard instances? There
are two aspects of the problem. The first is achieving a significant
probability of generation of hard instances. The second is keeping the
probability of easy instances negligible. A number of reductions (with
various limitations) exist between these tasks. Let us restrict our
attention to the first one.

First, note that Lemma~\ref{rnd} enumerates all distributions computable
in time $t(x)$, preserving $t$ within a linear factor. Thus, we can
generate the largest of all these distributions by adding them up with
summable coefficients, say, $1/i^2$. This distribution will be complete
for TIME$(t(x))$ and belong to TIME$(t(x)\|x\|)$. A nice alternative
would be to combine all complexities by translating high times into
small probabilities, similarly to Section~\ref {las}. Instead, we will
switch to samplable distributions directly.

\begin{definition}
Distributions generated by algorithms in L without input are called
\emph{samplable}. If the algorithm has inputs, they are treated as a
parameter for a \emph{family of samplable distributions}.
\end{definition}

Usual versions of this definitions are broader, allowing a class of
algorithms closer to LV(P) and generation of polynomially larger
distributions; we limit the generators to L for greater precision.

\begin{proposition}\label{opt}
There exists a complete, i.e., largest up to constant factors, family of
samplable distributions. \end{proposition}

The lemma follows if we note that L is enumerable and, thus, the
complete distribution can be obtained by choosing members of L at random
and running them. The generator of this distribution spends $O(1)$
average time per run and has the greatest in L (up to a constant factor)
capacity for generating surprises. (Compared to LV(P) algorithms, its
chance of a nasty surprise may be polynomially smaller.)

Since the complete distribution makes sense only to within a constant
factor, it is robust in logarithmic scale only and defines an objective
hardness of ``hitting'' a set $X$ given $x$.

\begin{notation}
By $\mathop{\rm Kl}(X/x)$, we denote the $-\log$ of probability of a set
$X$ under the complete distribution family parameterized by $x$.
\end{notation}

As often happens, a strong attack tool helps inventing a strong defense.
Optimal searches were noted, e.g., in \cite{search,laws,bnt} but here we
get a nice version for free. The complete generator of hard instances
can be used as an optimal algorithm for solving them. Since algorithms
in L combine time into probability, their performance is measured by
their chance of success. The generator of an optimal distribution family
(parameterized by the instance $x$) has the highest, up to a constant
factor, chance $1/S(f/x)=2^{-\mathop{\rm Kl}(f^{-1}(x)/x)}$ of
generating solutions. Its minus logarithm $s(f/x)$ measures the hardness
of each individual instance $x$ and is called its \emph{security}. The
generator takes an $O(1)$ average time per run and succeeds in expected
$S(f/x)$ runs. No other method can~do~better.

\paragraph{Open problem.}
The constant factor in the optimal inversion algorithm may be
arbitrarily large. It is unknown whether it can be limited to a fixed
constant (say, 10) independent of the competing algorithm for
sufficiently large instances.

\subsection{Inversion Problems and OWFs}

A complete distribution achieves about as high probability of hard
instances as possible. Using it makes the choice of a hard-to-invert
function easy: all NP-complete functions would do similarly well.
However, the interesting goal is usually to find a function that is
tough for some standard, say, uniform distribution of instances.
Serendipitously, Lemma~\ref{rnd} transforms any P-time distribution into
a uniform one via an appropriate encoding. Combined with this encoding,
any NP-complete function joins those hardest to invert. However, the
encoding destroys the function's appeal, so the problem of combining a
nice function with a nice distribution remains.

Of course, while many functions seem hard, none are proven to be such.
In \cite{rp,samp,color,G90,VR,JW}, a~number of combinatorial and
algebraic problems are proved to be complete on average with uniform
distributions, i.e., as hard as any inversion problem with samplable
distribution could be.

These results, however, do not quite yield owfs. The difference between
owf and complete on average inversion problems can be described in many
ways. The simplest one is to define owfs as hard on average problems of
inverting \emph{length-preserving} functions. In this case the choice
between picking at random a witness (crucial for owfs), or an instance,
becomes unimportant.

Indeed, each witness gives only one instance. So if length is preserved,
the uniform distribution restricted to solvable instances does not
exceed the one generated by uniformly distributed witnesses. On the
other hand, if the witnesses are mapped into much fewer hard instances,
they must have many siblings. Then, the function can be modified as
follows. Guess the logarithm $k$ of the number of siblings of the
witness $w$ and pick a random member $a$ of a universal hash family
$h_a(w)$. Output $f(w),k,a,h'_a(w)$, where $h'$ is $h$ truncated to $k$
bits. (These outputs, i.e. instances, are $k$ bits too long but can be
hashed into strings of the same length as the witnesses.) The extra
information in the output (if $k$ is guessed correctly) is nearly
random, and so makes inversion no easier. However, the siblings are
separated into small groups and the numbers (and, thus, uniform
probabilities) of hard instances and their witnesses become comparable.
The converse is also true:

\begin{proposition} Any owf with multimedian $V(k)$ of its security
$S(x)$ (for uniformly random $w$ and $x=f(w)$) can be transformed into a
length-preserving owf with a $1/O(k)$ fraction of instances that have
security polynomially related to $V$. \end{proposition}

First, the fraction of hard instances is boosted as described at the end
of Section~\ref{mult}. If the number of hard instances is much smaller
than that of their witnesses, the function can still be made
length-preserving without altering its hardness by separating the
siblings as in the previous paragraph. See \cite{samp} for more detailed
computations of the results of hashing owfs.

\subsection{Complete OWF: Tiling Expansion}

We now modify the Tiling Problem to create a complete combinatorial owf.
No such owf has been described yet, though \cite{ow} shows the existence
of an artificial complete owf. It would be nice to have several complete
owfs that are less artificial, i.e., intuitive to someone who does not
care to know the definition of computability. Below, such an example is
given as a seed. Hopefully, a~critical mass of such examples will be
achieved some day providing an arsenal for reductions to more popular
owf candidates to show their completeness.

\paragraph{NP versus owf completeness.}
It is a mystery why the industry of proving worst-case NP-completeness
of nice combinatorial problems is so successful. Of course, such
questions refer to art rather than science and so need not have a
definite answer. It seems important that several ``clean'' complete
combinatorial NP problems are readily available without awkward
complexities that plague \emph{deterministic} computation models. Yet,
\emph{average-case} completeness is only slowly overcoming this barrier.
And for complete \emph{owfs}, my appeal to produce even one clean
example remained unanswered for two decades.

A complete owf is easy to construct as a modification of a universal
Turing Machine (UTM). UTM computation, in turn, is easy to transform
into combinatorial objects (tiled squares, etc.). These objects are
stripped from the many technicalities needed to define a model of
computation. Indeed, these technicalities are involved in making
computation deterministic, which is unnecessary since the final relation
is intended to be nondeterministic anyway. The ``stripped'' versions
have simple combinatorial structure and appeal, which makes it possible
to find so many reductions to other mathematical objects.

This approach works for constructing average-complete NP problems. It
fails, however, to preserve length which is needed for constructing
complete owfs. (In some other definitions, length-preservation is
replaced with other requirements, but these, too, are destroyed by the
above construction.) Here, I use simple tools, such as the concept of
expansion, to try to make a first step to overcome these problems. The
construction preserves input length and retains clean combinatorial
structure of tiling. I hope that this first step could be used as a
master-problem for reductions to other nice owfs.

 \def\tilet {Tiles: unit squares with a letter at each corner;\\
             may be joined if the letters match.}
 \def\expan {Expansion: maximal tile-by-tile {\bf unique} (using given
tiles)\\ extension of a partial tiling of a square with marked border.}

\newlength\tilel\setlength\tilel{\columnwidth}
\newlength\tbx\setlength\tbx{10ex}
\addtolength\tilel{-\tbx}\addtolength\tilel{-1em}
\medskip\noindent\fbox{\parbox{\tilel}{\tilet\\ \expan}
\parbox{\tbx}{\mbox
 {\fbox{\parbox{3ex} {\makebox[3ex]{a\hfill x} \makebox[3ex]{e\hfill
	 r}}}
  \fbox{\parbox{3ex} {\makebox[3ex]{x\hfill c} \makebox[3ex]{r\hfill
	z}}}}
  \par\vspace {.5ex} \mbox
 {\fbox{\parbox{3ex} {\makebox[3ex]{e\hfill r} \makebox[3ex]{n\hfill
	 s}}}
  \fbox{\parbox{3ex} {\makebox[3ex]{r\hfill z} \makebox[3ex]{s\hfill
	z}}}}}}

\begin{definition}
\emph{Tiling Expansion} is the following function: Expand a given top
line of tiles to a square using a given set of \emph{permitted tiles};
output the bottom line and the permitted tiles. \end{definition}

\begin{theorem}
Tiling Expansion is a owf if and only if owfs exist. \end{theorem}

\paragraph{The reduction.}
We start with a UTM, add a time counter that aborts after, say, $n^2$
steps. We~preserve a copy of the program prefix and force it, as well as
the input length, on the output. This produces a complete
``computational'' length-preserving owf. Now, we reduce the computation
of the UTM, modified as above, to Tiling in a standard way. We add a
special \emph{border symbol} and restrict the tiles so that it can
combine only with the input or output alphabets (of equal size), or~with
the \emph{end-tape} symbol, or the state initiating the computation,
depending on the sides of the tile. The expansion concept does the rest.

Tiling is a nice combinatorial entity, but it lacks determinism. This
demands specifying all tiles in the square, which ruins the length
preservation. Requiring the set of tiles to force deterministic
computation would involve awkward definitions unlikely to inspire
connections with noncomputational problems, reductions to which is the
ultimate goal. Instead, expansion allows an arbitrary set of tiles but
permits the extension of a partially tiled square only at places where
such possible extension is unique. In this way, some partially tiled
squares can be completed one tile at a time; others get stuck. This
process is inefficient, but efficiency loss (small in parallel models)
is not crucial here.

It remains an interesting problem to reduce this owf to other nice
combinatorial or algebraic owfs thus proving their completeness.

The raw owfs, however, are hard to use. While many results, such as,
e.g., pseudo-random generators, require no other assumptions, their
constructions destroy efficiency almost entirely. To~be useful, owfs
need other properties, e.g., low Renyi entropy. This low entropy is also
required for transforming weak owfs into strong ones (while preserving
security, as in \cite{expnd}), and for other tasks. The following note
suggests a way that may achieve this. (Its $f(x)+ax$ can be replaced
with other hashings.)

\begin{remark}
Inputs of $g(a,x)=(a,f(x)+ax)$ have $\le1$ siblings on average for any
length-preserving~$f$ and $a,x\in$GF$_{2^{\|x\|}}$. \end{remark}

\begin{conjecture}
The above $g$ is one-way, for any owf $f$, and has the same (within a
polynomial factor) security. \end{conjecture}

\end{document}


\sloppy
 \newtheorem {definition} {\rm О~п~р~е~д~е~л~е~н~и~е~}
 \newtheorem{proposition} {\rm П~р~е~д~л~о~ж~е~н~и~е~}

 \newtheorem{lemma*}{\rm Л~е~м~м~а~}
 \newtheorem{theorem*}{\rm Т~е~о~р~е~м~а~}
 \date {Проблемы Передачи Информации 39(1), 2003\\
 (= Problems of Information Transmission) UDK 621.391:519.2}

 \newcommand\T{{\mathbb T}} \newcommand\R{{\mathbb R}}
 \newcommand\Z{{\mathbb Z}} \newcommand\Q{{Q}}\newcommand\N{{\mathbb N}}
 \newcommand\E{{\mbox{\bf E}}} \newcommand\e\varepsilon
 \renewcommand\C{{\mathbb C}} \renewcommand\a\alpha

\author{Леoнид~А.~Левин}\title{ОДНОСТОРОННИЕ ФУНКЦИИ\thanks
 {Работа выполнена при частичной финансовой поддержке Clay Mathematics
   Institute, Boston University, IHES и NSF (Grant CCR-9820934).}}

 \maketitle\renewcommand{\abstractname}{\strut}
 \begin{flushright}\vskip 2\baselineskip\parbox{19pc}{\begin{tabbing}
 \emph {Слаб император и вся его рать}\\*
 \emph {Снова Шалтая-Болтая собрать.}\footnotemark
 \end{tabbing}} \end{flushright} \footnotetext
 {Эти строчки английской загадки комбинируют хорошо известный
 читателю перевод Маршака со следующим более буквальным переводом:
 {Круглик-Горбик на стенке сидел, /$\!$/
 Круглик-Горбик с треском слетел. /$\!$/
 Слаб император, слаба его рать /$\!$/
 Круглика-Горбика снова собрать}. ---~\emph{Примечание Л.А. Левина}.}

\begin{abstract}\vskip 1\baselineskip
 Одной из наиболее важных проблем теоретической информатики является
вопрос о существовании односторонних функций. Эта статья рассматривает и
уточняет ряд понятий, с ним связанных. В частности, впервые приводится
явное комбинаторное построение полной односторонней функции (полнота
означает, что эта функция является односторонней, если таковые вообще
существуют). Основные концепции содержат неожиданно много тонкостей
(частично уже упоминавшихся в литературе). Здесь предлагается некоторый
единый подход. \end{abstract}

\section{Введение I: обращение функций}

С незапамятных времен человечеству не раз напоминалось~-- иногда
жестоко~-- что сделанного не обратишь. Математики столкнулись с
формальным выражением этой проблемы, едва начав серьезно анализировать
основания своей науки.

\textbf{1.1. Странная аксиома.} Более века назад Георг Кантор предложил
свести все разнообразие математических понятий к единственному понятию
множества, а все математические теоремы~-- к единственной схеме аксиом,
которую можно назвать \emph{постулатом Кантора}. Этот постулат
утверждает (для каждой формулы $A(x)$ в языке теории множеств), что
существует множество, которое состоит из всех $x$, для которых выполнено
$A(x)$. Это безобидное утверждение (почти что определение понятия
множества), как вскоре выяснилось, имеет много следствий, и даже больше,
чем хотелось бы: из него можно вывести противоречие. Чтобы спасти
положение, Цермело, Френкель и другие математики прагматически
ограничили постулат Кантора некоторыми его частными случаями, разрешив
лишь формулы $A$ специального вида. При этом вроде бы противоречия не
получается, а все ценное продолжает выводиться. Получилась
аксиоматическая теория множеств, играющая центральную роль в основаниях
математики.

В 1904 году Цермело заметил, что в доказательствах математических теорем
используется еще одна аксиома, (печально) знаменитая \emph{аксиома
выбора}. Ее можно сформулировать так: для всякой функции $f$ существует
обратная к ней, {т.е.} такая функция $g$, для которой $f(g(x))=x$ при
всех $x$ из области значений функции $f$. Математики постепенно приняли
ее, хотя и неохотно: до сих пор использование этой аксиомы отмечается
особо. Аксиома выбора не является частным случаем постулата Кантора, к
тому же имеет парадоксальные следствия. Вот простой пример, основанный
на том, что эта аксиома позволяет \emph{симметричный} выбор
произвольного целого числа.

Рассмотрим аддитивную группу $\T=\R/\Z$ вещественных чисел по модулю
$1$, естественно отождествляемую с окружностью и с промежутком $[0,1)$.
Пусть $\Q_{10}$~-- ее подгруппа, состоящая из всех конечных десятичных
дробей, {т.е.} чисел вида $a/10^b$. Пусть $f$ сопоставляет с каждым
числом $x\in\T$ его смежный класс $x+\Q_{10}$ в фактор-группе
$\T/\Q_{10}$.

Любая обратная к $f$ функция $g$ выбирает в каждом смежном классе по
представителю. Пусть $G=g(f(\T))$~-- множество этих представителей,
{т.е.} множество значений функции $g$. В этом случае для каждого $x\in
\T$ существует единственный сдвиг из $\Q_{10}$, переводящий $x$ внутрь
$G$ (и $\T$ разбивается на счетное число классов, получающихся из $G$
сдвигом на элементы $\Q_{10}$).

Мы несколько отступим от традиционного изложения, чтобы подчеркнуть
элементарный характер рассматриваемого парадокса. Еще одно (последнее)
обозначение: для любого числа $q\in \Q_{10}$ через $q'$ мы обозначим
число $(10 q)\bmod 1$, получаемое отбрасыванием старшего разряда.

Пусть $p,q$~-- элементы $\Q_{10}$. Я предлагаю такое пари: если $x+p\in
G$ (для случайно выбранной точки $x$ на окружности), то вы платите мне
рубль, а если $x+q\in G$, то я плачу вам два. Это выгодно для вас,
поскольку моя ставка больше, а условия выигрыша совершенно симметричны
(соответствующие множества отличаются поворотом). Мало того, я готов
быть еще более щедрым и одновременно заключить много таких пари, по
одному для каждого $q\in \Q_{10}, q>0$; я ставлю на $p=q'$ про\-тив~$q$.
Как только вы соглашаетесь, мы выбираем случайное $x\in\T$ (каждая его
цифра получается бросанием честной монеты) и подсчитываем наши прибыли и
убытки. Пусть $\boldsymbol{q}$~-- тот единственный элемент множества
$\Q_{10}$, для которого $x+\boldsymbol{q}\in G$. Тогда я проигрываю пари
для этого $\boldsymbol{q}$ и выигрываю десять пари (для всех $q$, при
которых $q'=\boldsymbol{q}$). Щедрость окупается, не правда ли?

Не так просто объяснить, в чем ``корень зла''\ в этом парадоксе. В
каждой игре срабатывают лишь $11$ пари, и о ``финансовой пирамиде''\
речь не идет. Более того, если брать в качестве $x$ случайную точку
сферы, то даже общее число всех возможных пари можно сделать конечным:
Банах и Тарский \cite{1} показали, что можно построить шесть пар $(A_i,
T_i)$, где $A_i$ есть некоторое подмножество сферы $S^2$, а $T_i$~--
поворот этой сферы, с таким свойством: ставя на $x\in A_i$ против
$T_i(x)\in A_i$ одновременно при всех $i=1,\ldots,6$, мы проигрываем
одно пари и выигрываем два (при любом $x$). Возвращаясь к нашему
примеру, отметим, что числа $x+p$ и $x+q$ проверяются на принадлежность
к одному и тому же множеству $G$ и отличаются в конечном числе
десятичных разрядов (причем все разряды независимы и равновероятны).

Позволительно усомниться вообще в законности мысленного эксперимента,
использующего бесконечное число случайных бросаний, или в осмысленности
вопроса о принадлежности $x+q$ множеству $G$. Но при этом мы покушаемся
на самые основы теории множеств. Проще интерпретировать отказ от
кажущегося выгодным пари как скрытое недоверие к аксиоме выбора.

\textbf{1.2. Конечные объекты и полный перебор.} Теорию вычислений эти
трудности с обращением функций затрагивают меньше, поскольку конечные
объекты и так вполне упорядочены (принципом математической индукции) и
аксиома выбора ни к чему. (Тем не менее вопрос об \emph{эффективном
обращении} функции остается актуальным.) К тому же шенноновская теория
информации для любой функции $f$ приписывает случайной величине $x$
столько же информации о случайной величине $f(x)$, сколько $f(x)$ об
$x$. Колмогоровская (алгоритмическая) теория информации~(см.~\cite{2,3})
позволяет определить понятие количества информации для конечных (не
обязательно случайных) объектов: $I(x:y)$, количество информации в $x$
об $y$, есть разность между длинами кратчайших программ, порождающих $y$
без использования $x$ и с использованием $x$. Колмогоров и автор
настоящей статьи доказали в 1967~году, что эта величина, подобно
шенноновской, симметрична: $I(x:y)\approx I(y:x)$, хотя это равенство
лишь приближенное~\cite{4}.

В доказательстве есть подвох: используется полный перебор всех строк
данной длины, требующий невообразимо огромного времени. Например,
произведение двух простых чисел содержит всю информацию об этих числах
(и наоборот), но извлечь эту информацию на практике невозможно~-- именно
на этом предположении основана система RSA~\cite{5} и множество других
приемов современной криптографии. Колмогоров в свое время указывал на
свойство симметрии информации как на одну из задач, хорошо подходящих
для попытки доказать, что полный перебор неустраним
($\mathrm{P}\ne\mathrm{NP}$, как сказали бы сейчас) \cite{6}.

Появление системы RSA изменило взгляд на трудно обратимые функции:
досадное препятствие стало незаменимым инструментом. За RSA последовало
множество других удивительных приложений. Оказалось (и это существенно
для многих из них), что трудность обращения функции можно
сконцентрировать в единственном бите. Говорят, что (легко вычислимый)
предикат $b(x)$ является \emph{трудным битом} (hard-core bit) для трудно
обратимой (односторонней, one-way) функции $f(x)$, если восстановить
значение $b(x)$ по $f(x)$ (или угадать со сколько-нибудь заметной
корреляцией) столь же сложно, как восстановить все $x$.

Впервые такой предикат был предложен в \cite{7} для функции $f(x)=
a^x\bmod p$. Вскоре были указаны трудные биты для системы RSA, для
функции Рабина ($x\mapsto\linebreak \mapsto x^2\bmod n$, где $n$ есть
произведение двух простых чисел), а также для ``дробящих''\ функций вида
$f^*(x_1,\ldots,x_n)= f(x_1),\ldots, f(x_n)$ (см.~\cite{8};
доказательство леммы об изоляции, используемой в~\cite{8} при анализе
трудности бита, можно найти в~\cite{9}). В~\cite{10} решен общий случай
произвольной односторонней функции (см. также~\cite {L93,Knuth}).

Почему трудные биты так важны? В \cite{7, 8} показано, как с их помощью
можно преобразовать короткое случайное двоичное слово (seed) в
неограниченную последовательность битов, неотличимых от случайных. Если
односторонняя функция $f$ является перестановкой, то годится
последовательность $g_s(i)=b(f^i(s))$, $i=0,1,2,\ldots$ Общий случай
разработан в~\cite{11}.

Таким образом барьер между случайными и детерминированными процессами
был размыт, и в 1980-х годах были получены многие результаты, казавшиеся
ранее немыслимыми. Среди них общие теоремы криптографии (см., например,
\cite{12}), конструкции доказательств с нулевым разглашением
(zero-knowledge proofs) и реализация произвольных протоколов
взаимодействия не доверяющих другу участников (как бывает, скажем, в
карточных играх) в виде игр с полной информацией \cite{13}. Этот период
можно назвать золотым веком теоретической информатики, и он стал
возможным благодаря односторонним функциям.

\section{Введение II:\\ экстравагантные вычислительные модели}

\textbf{2.1. Падение RSA.} Заметим, что все перечисленные конструкции
основаны на односторонних (легко вычислимых и трудно обратимых)
функциях, а существование таких функций остается лишь гипотезой,
неоднократно подвергавшейся разного рода сомнениям.

Первый эпизод такого рода связан с самим Шамиром, одним из изобретателей
системы RSA. Он доказал в \cite{14}, что разложение на множители (на
трудности которого основана RSA) может быть выполнено за полиномиальное
число арифметических действий. При этом каждое действие считается за
одну операцию, независимо от длины чисел (``unit-cost model''). При
возведении в квадрат длина числа удваивается, и повторное возведение в
квадрат быстро приводит к числам космологического размера. Одно такое
число может кодировать длинный массив обычных чисел, и одной операции
соответствует большой объем работы (например, проверка экспоненциального
числа возможных делителей). Алгоритм Шамира не произвел впечатления на
косных криптографов: отвергнутой оказалась вычислительная модель,
игнорирующая длину чисел, а не RSA.

Другая не лишенная сходства атака на RSA происходит в настоящее время и
началась с замечательного результата Питера Шора. Он показал, что можно
разлагать числа на множители за полиномиальное время, используя
воображаемые аналоговые вычислительные машины, названные
\emph{квантовыми компьютерами}. Эти машины подсказаны законами квантовой
физики (доведенными до крайности).

\textbf{2.2. Квантовые компьютеры.} Квантовый компьютер состоит из $n$
взаимодействующих элементов, называемых \emph{кубитами} (q-bits). Каждый
из них является квантовой системой; ее чистые состояния представляют
собой единичные векторы в двумерном комплексном пространстве $\C^2$; две
компоненты вектора представляют собой амплитуды двух булевских значений.
Состояние системы в целом представляет собой вектор в тензорном
произведении $n$ двумерных пространств. В этом произведении есть базис
из $2^n$ векторов, являющихся произведениями базисных векторов
сомножителей. Каждый базисный вектор, таким образом, соответствует
двоичному слову длины $n$. Квантовый компьютер охлаждают, изолируют от
внешнего мира почти идеально и помещают в базисное состояние, в котором
часть битов образует входное слово, а остальные биты равны нулю. Далее
производится вычисление, состоящее из последовательного применения
идеально обратимых преобразований, соответствующих взаимодействию
кубитов. Полученное состояние может быть суперпозицией невообразимо
быстро растущего числа базисных состояний с экспоненциально малыми
амплитудами. Взаимодействие с окружающей средой может вносить ошибки,
поэтому вычисление предусматривает коррекцию этих ошибок (возможную,
если ошибки редки и имеют специальный вид). В остальном мы полагаемся на
законы квантовой механики, считая их выполненными с неограниченной
точностью. Последнее предположение весьма существенно, поскольку
амплитуды экспоненциально малы и отклонения в сотом (или даже намного
более далеком) знаке после запятой могут полностью изменить ход
вычисления.

В \cite{15} показано, что такие компьютеры могут выполнять разложение на
множители за полиномиальное время. Используя грубую аналогию, можно
сказать, что каждая из экспоненциального числа координат проверяет
делимость на соответствующее число, и амплитуды концентрируются в тех
координатах, которые соответствуют делителям.

\textbf{2.3. Маленькие трудности.} Попытка реализовать компьютеры
описанного вида сталкивается с массой проблем. Скажем, трудно
представить себе идеальную термоизоляцию, не говоря уже о защите от
нейтрино, гравитационных волн и тому подобной экзотики; их влияние на
квантовые амплитуды не обязательно удовлетворяет условиям, от которых
зависит коррекция ошибок. Более того, даже и классические вычисления без
рассеяния тепла остаются гипотетическими, хотя про них говорят уже
несколько десятилетий, а в воображаемых мирах с искусственными законами
взаимодействия их существование даже доказано. В реальном мире, где нам
доступно в основном лишь электромагнитное взаимодействие между
электронами, ядрами и фотонами, такого рода схемы (производящие большие
вычисления с малым тепловыделением) остаются спекулятивными. Поэтому
понижение температур имеет предел, а даже и небольшое количество тепла
может сильно нарушить когерентность. Более того, неконтролируемые
степени свободы могут быть более опасны, чем простое нагревание: кто
знает, к чему может привести их взаимодействие с тончайше
скоррелированными состояниями кубитов?

\textbf{2.4. За сотни цифр от запятой.} Все это цветочки. Основная
проблема другая: в требовании, чтобы законы квантовой механики были
справедливы с той фантастической точностью (сотни, если не миллионы
знаков после запятой), которая нужна для квантовых алгоритмов. Физики не
знают ни одного закона, который был бы справедлив с точностью до
нескольких десятков знаков. Как учит история физики, каждые несколько
следующих знаков требуют новых теорий и переосмысления базисных понятий.
Может, при таком уровне точности надо считать квантовые амплитуды не
комплексными числами, а кватернионами, раскрашенными графами или
гремлинами с черным юмором\ldots\ Я подозреваю, что физики и в законы
арифметики-то с такой точностью не верят. (На самом деле мы даже знаем,
что основные законы физики не могут выполняться с такой точностью,
поскольку они начинают противоречить друг другу!)

И вообще, какой может быть физический смысл в $500$-значном числе? Пусть
мы говорим: ``В этом ящике хранятся кубиты, которые содержат текст
десяти заповедей, с амплитудой, в которой три десятичные цифры, начиная
с пятисотой, равны $666$''. Есть ли хоть какой-то физический смысл в
нашем утверждении? Если состояние системы близко к базисным векторам
тензорного произведения, то можно надеяться переформулировать
утверждение, заменив длинную десятичную дробь на несколько коротких
(которые уже можно измерять). Что-то осмысленное можно вообразить для
больших систем типа лазеров или конденсированных состояний, где
отдельными состояниями можно пренебречь. Но разложение на множители с
помощью квантовых компьютеров существенно использует экспоненциальное
число глубоко индивидуальных состояний. Мне трудно представить, чем в
такой ситуации можно заменить общее и прямое представление амплитуд с
невообразимой точностью, разве что сделав их более ``физическими''\ за
счет выбора менее физического базиса. Рассмотрим этот подход более
подробно.

\textbf{2.5. Как мала вселенная.} Сторонники квантовых компьютеров часто
говорят, что так или иначе мы будем в выигрыше: удастся либо сделать
работающий квантовый компьютер, либо уточнить законы квантовой механики.
Например, П. Шор в \cite{16} говорит: \begin{quotation}\noindent
 Если в квантовой механике имеются нелинейные эффекты, которые
 мож\-но обнаружить, наблюдая за неисправной работой квантовых
 компьютеров, физики ВЕСЬМА заинтересуются ими (я бы ожидал
 Нобелевской премии за убедительное свидетельство таких эффектов).
\end{quotation}

\medskip

Рассмотрим, однако, другой вариант: нам удается сделать квантовые
компьютеры из нескольких кубитов, работающие в соответствии с теорией. С
ростом числа кубитов, однако, такие компьютеры сделать все сложнее, и
прогресс останавливается задолго до того, как квантовые вычисления
способны соперничать с вычислениями на обороте старого конверта. Поэтому
разлагать больш\'ие числа на множители мы не можем. Можно ли хотя бы
претендовать на более или менее достойную премию за поправки к квантовой
механике?

Однако награждающий комитет захочет увидеть что-нибудь более конкретное,
чем просто неработающий компьютер,~-- например, явное указание
состояния, в котором он находится. Но всех ресурсов вселенной
оказывается недостаточно, чтобы измерить необходимое число координат с
необходимой точностью. (Остается лишь надеяться на Нобелевскую премию по
экономике, если удастся собрать деньги, необходимые для разложения
пятизначных чисел на множители с помощью квантовых компьютеров!)

Сделаем некоторые оценки, обозначая через ${\sim}n$ небольшие степени
числа $n$.\linebreak У криптографов числа, подлежащие разложению на
множители, могут иметь тысячи (а легко возможны и миллионы) битов.
Пространство состояний~$H$ размерности~$2^{\sim n}$ имеет примерно
$2^{2^{\sim n}}$ \emph{почти} ортогональных друг другу векторов.
Рассмотрим элемент~$v\in H$ общего положения (будем называть такие
векторы ``мегасостояниями''). Минимальный размер машины, которая может
породить или распознать~$v$, имеет порядок~$K=2^{\sim n}$~-- значительно
больше, чем вселенная. В самом деле, достаточно посчитать машины
из~$K$~атомов: их примерно~$2^{\sim K}$.

Есть принципиальная разницами между не наблюдавшимися и принципиально
ненаблюдаемыми вещами. Утверждения про отдельные мегасостояния
ненаблюдаемы. Можно представить себе способ отличить два мегасостояния
\emph{друг от друга}, но~-- как показывают сделанные оценки~-- нет
возможности отделить данное состояние (общего положения) от всех далеких
от него с той степенью точности, которая существенна для поведения
квантового компьютера. Какой, в таком случае, эксперимент может
подтвердить правильное состояние квантового компьютера? (Даже
мысленный~-- в предположении, что мы пользуемся ресурсами всей
вселенной, \emph{но не более}!)

Еще Архимед открыл, что цифровая форма представления чисел
экспоненциально эффективнее аналоговой (кучи песка). Впоследствии
различные аналоговые устройства редко оказывались впечатляющими. Неясно,
почему квантовые компьютеры должны быть исключением.

\textbf{2.6. Метрика или топология?} Говоря о приближениях к состояниям
системы, мы обнаруживаем, что в квантовой механике отсутствует
адекватный формализм. Есть тонкое различие между приближениями в смысле
метрики и топологии. Метрика говорит о том, насколько одно (хорошее)
состояние близко к специфическому другому (плохому). Топология имеет
дело с близостью хорошего состояния к сочетанию всех неприемлемых (``не
окрестных'') состояний, и это не обязательно то же самое, что расстояние
до ближайшего плохого состояния, особенно для квантовых систем.

В бесконечномерных пространствах есть разные способы отличать конечную
разделенность точки и множества от нулевой (разные топологии). В
конечномерном случае с формальной точки зрения такого различия нет: все
топологии совпадают. Однако числа порядка $2^{500}$ можно считать
конечными лишь в весьма философском смысле, и было бы желательно ввести
некое понятие слаботопологической \emph{глубины} окрестности,
полиномиально связанное с ресурсами, необходимыми для достижения этой
глубины. При этом точность, соответствующая разумной глубине, должна
быть физически достижимой (возможно порождать точки внутри
соответствующей окрестности, отделять центр окрестности от точек, в ней
не лежащих, и т.д.).

Имея функцию расстояния, мы можем говорить об $\varepsilon$-окрестностях
(что невозможно в топологии, где конкретное значение $\varepsilon$
отсутствует, известно лишь существование некоторого $\varepsilon>0$).
Однако $\varepsilon$-окрестности в метрических пространствах неизбежно
обладают таким свойством (которое можно назвать ``аксиомой
пересечения''): пересечение любого множества $\varepsilon$-окрестностей
(при данном $\varepsilon$) есть $\varepsilon$-окрестность. Говоря о
близости в пространстве состояний квантовой системы, мы может захотеть
измерять глубину окрестности числом, не подчиняясь аксиоме пересечения.
Вот пример такого рода измерения (не претендующий на пригодность для
наших целей). Предположим, что окрестность нуля задается системой
линейных неравенств $f_i(x)<1$; тогда ее глубиной считается число
$1/\sum\limits_i\|f_i\|$, хотя ограничение точки $x$ сферой с
гильбертовой нормой $1$ сделало бы эту глубину квадратично связанной с
радиусом окрестности.

Возможно, что более осмысленный с точки зрения физики подход можно
получить, если учитывать наличие специального базиса (составленного из
тензорных произведений).

\textbf{2.7. Не продешевить!} Разложение на множители с помощью
квантовых компьютеров кажется мне научной фантастикой. Можно лишь
сожалеть, что в популярных обзорах оно не отделяется от более
реалистичных предложений~-- квантовой криптографии, квантовой
информации, а также квантовых вычислений, в которых существенен
нелокальный доступ к информации (например, быстрый поиск в больших
массивах).

Следует отметить также, что фундаментальные причины, препятствующие
работе квантовых компьютеров, определенно не ясны и заслуживают
серьезного изучения. Польза от него может быть больше, чем вся мыслимая
польза от быстрого разложения на множители. Представьте себе, что
знаменитый демон Максвелла в свое время рекламировался бы как
перспективный способ получения электроэнергии из окружающего тепла!
Вероятно, в таком случае современная термодинамика появилась бы заметно
позже.

В оставшейся части статьи мы не рассматриваем экстравагантные модели
вычислений и твердо придерживаемся ``полиномиального тезиса
Черча--Тьюринга'': любое вычисление, требующее $t$ тактов работы
$s$-битового устройства, можно моделировать на машине Тьюринга за
$s^{O(1)}t$ шагов с памятью $s^{O(1)}$.

\section{Подвохи усреднения}

Сама по себе трудность обращения функции в худшем случае (на самом
трудном входе) еще мало что значит. Представим себе, например, что все
входы делятся на ``легкие''\ и ``трудные''. Легкие входы $x$ требуют
времени $\|x\|^2$ для обработки; а трудные~-- экспоненциального среднего
времени \emph{как для обработки, так и для нахождения} вероятностными
алгоритмами. В таком случае вселенная слишком мала, чтобы породить
пример, который мы не можем решить, и обращение функций не представляет
трудностей на практике.

На практике важна трудность обращения в ``типичных случаях''. Именно она
мешает разработчикам алгоритмов и делает возможными разные
криптографические чудеса. Но корректное определение ``типичности''~--
вещь довольно тонкая.

\textbf{3.1. Las Vegas-алгоритмы.} Прежде всего надо договориться о
способах измерения ``эффективности''\ алгоритма обращения. Алгоритм
обращения $A(x,\a)$ получает на вход значение $x=f(w)$ обращаемой
функции $f$ на некотором искомом $w$, а также последовательность
случайных битов $\a\in\{0,1\}^{\N}$. Если речь идет об обращении, когда
мы можем проверить ответ подстановкой, нам нет необходимости заботиться
о неверных ответах, и можно предполагать, что алгоритм либо дает верный
ответ (один из прообразов элемента $x$), либо не дает никакого ответа
(возможно, никогда не останавливается). Такие алгоритмы называют Las
Vegas-алгоритмами.

Эффективность работы алгоритма на данном примере $x$ измеряется двумя
параметрами: \emph{объемом} $V$ вычисления\footnote[3]
 {Мы говорим ``объем'', а не ``время'', имея в виду возможные модели
вычислений с неограниченным параллелизмом.}
 и \emph{вероятностью} $p_V$ (относительно $\a$) успешного обращения за
$V$ шагов. Эти параметры связаны друг с другом: можно увеличить
вероятность успеха $p$ за счет повторения $A$ при разных независимых
$\a$; при этом отношение $V/p$ (точнее $-V/\log(1\!-\!p)$) меняется
мало.
 Поэтому кажется разумным (как это часто и делается) фиксировать
значение $p$ и требовать, чтобы алгоритм выдавал верный ответ, скажем, с
вероятностью $1/2$ или больше. Однако при этом возникает трудность:
оценка вероятности $p$ и необходимого числа повторений может потребовать
экспоненциального объема (в худшем случае), и потому имеет смысл
говорить лишь о среднем объеме вычислений (для фиксированной вероятности
успеха). Подчеркнем, что усреднение выполняется лишь по внутренним
случайным битам ({т.е.} по $\a$), вход $x$ выбирается противником. При
таком подходе два параметра для измерения эффективности излишни:
увеличения вероятности можно достичь за счет увеличения среднего объема
и наоборот (чтобы уменьшить средний объем, достаточно выполнять
обращение лишь с некоторой малой вероятностью).

Можно пойти другим путем и нормализовать не вероятность успеха, а
средний объем вычислений. При этом мы избегаем накладных расходов на
оценку вероятности успеха и уменьшения модулярности из-за комбинирования
нескольких применений алгоритма. Есть, однако, более важные аргументы в
пользу такого подхода: представим себе, например, что различные
прообразы данного значения односторонней функции имеют разное (и трудно
сравнимое) ``качество''. Ограничения же среднего объема вычисления имеют
вполне ясный смысл. Для конкретной модели вычислений можно фиксировать
$O(1)$-границу для объема; если же мы хотим работать с различными
разумными моделями, можно ограничивать средний объем вычислений
многочленом от длины входа (при этом для разных алгоритмов можно брать
разные многочлены). Тут, однако, возникает препятствие: множество
алгоритмов с данным ограничением на средний объем вычислений не является
рекурсивно перечислимым. Преодолеть эту трудность можно с помощью
следующего подхода, гарантированно ограничивающего средний объем
вычислений.

\begin{definition}
\emph{Las Vegas}-алгоритм $A(x,\a)$ из класса $\mathrm{LV}(b)$ начинает
работу, зная заранее границу $b(x)$ на допустимый объем вычислений. При
этом в любой момент алгоритм имеет право поставить ``на кон''\ любую
часть оставшегося объема; если он проигрывает $($что определяется
следующим случайным битом $\a)$, то эта часть пропадает, если же он
выигрывает, то она удваивается. Обычно достаточно рассматривать класс
$\mathrm{LV}(O(1))$, который мы обозначаем просто
$\mathrm{L}$.\footnote[4]
 {Это ``L''\ можно произносить как ``Las'', имея в виду
 Las Vegas, а также Laszlo Babai (который придумал название
 ``Las Vegas-алгоритмы'').
 Подчеркнем, что речь не идет об
 определении очередного сложностного класса: мы описываем
 некоторую определенную \emph{форму алгоритмов}, а не класс
 алгоритмов или (что еще более абстрактно) класс задач,
 разрешимых с помощью алгоритмов некоторого класса. Отметим,
 что по существу требование $\mathrm{L}$ не является особенно новым:
 оно представляет собой лишь небольшое усиление общей
 идеи Las Vegas-алгоритма.}
 \end{definition}

Несмотря на жесткое $O(1)$-ограничение, $\mathrm{L}$-алгоритмы
достаточно пред\-ста\-ви\-те\-ль\-ны: любой Las Vegas-алгоритм можно
привести к виду из $\mathrm{L}$, сохраняя (в основном) отношение между
сложностью и вероятностью успеха. Если $p$~-- вероятность успеха для
$\mathrm{L}$-алгоритма, величина $1/p$ соответствует числу повторений,
необходимому для достижения успеха с фиксированной вероятностью, и
потому играет роль времени работы алгоритма. При этом читатель, не
желающий вдаваться в детали внутреннего устройства компьютеров, может
принять $\mathrm{L}$-алгоритмы как некую данность и весь дальнейший
анализ проводить исключительно в терминах теории вероятностей!

\textbf{3.2. Оценка времени в терминах мультимедиан.} Нам осталось
разобраться с более сложной проблемой~-- усреднением по возможным входам
$x$. Определение сложности как среднего значения $\E_x t(x)$ (для
алгоритма, делающего $t(x)$ шагов на входе $x$) весьма неустойчиво. В
самом деле, при переходе от одной вычислительной модели к другой время
работы может возвестись в квадрат. Такова ситуация, например, с
распознаванием симметрии входа: на одноленточной машине оно требует
квадратичного времени, а на двухленточной достаточно линейного.
Представим себе, что подобная ситуация имеет место для значительно более
медленных алгоритмов. Пусть, например, время $t(x)$ работы алгоритма
равно $\|x\|^2$ для всех входов, кроме состоящих из одних нулей, а для
$x=0^n$ равно $2^n$ в одной модели вычислений и $4^n$ в другой. Будем
считать, что все $2^n$ входов данной длины $n$ равновероятны. Тогда
среднее значение $\E_x t(x)$ будет полиномиальным в одной модели и
экспоненциальным в другой: усреднение не коммутирует с возведением в
квадрат. К тому же эта экспоненциальная оценка сложности на среднем
входе не имеет практической ценности, поскольку вероятность появления
сложного входа пренебрежимо мала.

Более инвариантной мерой сложности вычисления могла бы служить
\emph{медиана} времени вычислений на случайном входе, {т.е.} минимальное
число шагов, достаточное для обработки любого из \emph{более сложной
половины} входов. Эта мера, однако, неустойчива в другом смысле: она
может кардинально измениться, если \emph{половину} наиболее сложных
входов заменить, скажем, на \emph{четверть}.

К счастью, наше соглашение о Las Vegas-алгоритмах заодно решает и эту
проблему. Для произвольного $\mathrm{L}$-алгоритма мы можем измерить
вероятность успешного обращения на случайном входе (имеющем заданное
распределение вероятностей). Обратная величина к этой вероятности (как
функция, скажем, от размера входа) является разумной мерой
\emph{стойкости} односторонней функции. Такая мера хороша для
криптографии, где ставится цель предотвратить взлом шифра даже и с малой
вероятностью. В задачах, где требуется достичь успеха на почти всех
входах, необходим другой подход.

 \begin{definition}
Пусть задано некоторое $\mathrm{L}$-распределение на входах $($говоря о
$\mathrm{L}$-рас\-пре\-де\-ле\-ни\-ях, мы имеем в виду распределения на
выходах $\mathrm{L}$-алгоритма с пустым входом$)$\footnote[5]
 {Если распределение на входах не алгоритмическое, можно
 изменить определение, заменив в определении
 класса $\mathrm{L}$ сложность на длину выхода. При этом
 входы длины $n$ должны иметь вероятности с суммой $n^{-O(1)}$,
 например, $p(x)=1/(\|x\|\log\|x\|)^2$.}. Порождаем вход $($по этому
распределению$)$ $k$ раз, что требует среднего времени $O(k)$, а затем
применяем алгоритм обращения, пока не будут найдены прообразы у всех
входов, для которых они существуют\footnote[6]{Входы, для которых
прообразов нет, не учитываются.}. Число попыток является случайной
величиной $($зависящей от случайных битов, используемых в алгоритме
обращения, а также от случайных битов, использованных при порождении
входов$)$. Ее медиана $\mathrm{MT}(k)$ называется мультимедианой времени
обраще\-ния~$f$ с помощью алгоритма $A$.
 \end{definition}

Эта мера качества алгоритма имеет ряд достоинств. Она коммутирует с
возведением в квадрат времени работы и потому устойчива относительно
перехода к другой модели вычислений. Выбор границы $1/2$,
подразумеваемой медианой, также не является существенным. В самом деле,
увеличение $k$ в $c$ раз соответствует такому же увеличению
$\mathrm{MT}(k)$ как и уменьшение вероятности неудачи обращения до
$2^{-c}$.

Величина $\mathrm{MT}$ осмысленна и для верхних, и для нижних оценок.
Пусть $t(x)$ велико для $\varepsilon$-доли входов $x\in\{0,1\}^n$. Тогда
$\mathrm{MT}(k)$ столь же велико для $k=n^3/\e$. Обратно, пусть
$\mathrm{MT}(k)$ велико. Тогда почти наверняка $t(x_i)$ столь же велико
для некоторых из $n=k^2$ случайных входов $x_1,\ldots, x_n$ (и
$\sum\limits_i\|x_i\|=O(n)$).

\textbf{3.3. Простые распределения.} До сих пор мы учитывали влияние
случайных битов в вероятностном алгоритме на время работы (для
фиксированного входа), а также усредняли это время по различным входам с
фиксированным распределением вероятностей. Сейчас мы обсудим выбор этого
распределения, который далеко не всегда прост и отнюдь не исчерпывается
ссылкой на ``равномерное распределение''. Такие ссылки часто лишь
запутывают дело, поскольку различные распределения могут не без
оснований считаться ``равномерными''.

Например, рассмотрим графы $G=(V,E\!\subset\! V^2)$ с $n$ вершинами
($\|V\|=n$), где значение $n$ выбирается с вероятностью $c/n^{2}$ (здесь
$c$~-- нормирующий множитель). На этом множестве (при данном $n$)
рассмотрим два распределения, которые заслуживают названия
``равномерных''. Первое из них, которое мы обозначим $\mu_1$, случайно и
равновероятно выбирает граф $G$ среди всех $2^{n^2}$ графов.
Распределение $\mu_2$ соответствует случайному равномерному выбору числа
ребер $k=\|E\|$ в диапазоне от $0$ до $n^2$ и затем случайному
равномерному выбору $E$ среди $C^k_{n^2}$ возможностей. Хотя оба
распределения могут быть названы равномерными, они радикально
отличаются. Например, множество $\{G: \|E\|=n^{1{,}5}\}$ имеет
вероятность примерно $1/n^2$ относительно $\mu_2$, хотя его вероятность
относительно $\mu_1$ экспоненциально мала.

В определенном смысле все ``простые''\ распределения вероятностей можно
считать равномерными. Мы приведем соответствующее рассуждение (кратко
намеченное в \cite{17}) с некоторыми добавлениями, нужными для
дальнейшего.

Отождествим (как это делают в теории множеств) каждое натуральное
чис\-ло~$n$ с множеством меньших натуральных чисел
$\{0,1,2,\ldots,n-1\}$. Мерой мы будем называть аддитивную
вещественнозначную функцию на \emph{множествах} натуральных чисел. В
соответствии с нашим соглашением
$\mu(n)=\mu(\{0\})+\mu(\{1\})+\ldots+\mu(\{n-1\})$; тем самым $\mu$
представляет собой монотонную \emph{функцию распределения}.
Соответствующая \emph{плотность} распределения задается формулой
$\mu'(n)= \mu(n+1)-\mu(n)=\mu(\{n\})$; величина $\mu'(n)$ есть
вероятность одноэлементного множества $\{n\}$ (а не $n$ как множества
меньших натуральных чисел). Через $\Q_2$ обозначим множество конечных
двоичных дробей $i/2^{\|i\|}\in[1/2,1)$. Мы округляем значения функции
$\mu$ до элементов $\Q_2$, сохраняя лишь минимально необходимое число
двоичных цифр (так, чтобы вероятность изменилась не более чем в
константу раз).

\begin{definition}
 Говорят, что функция $\mu\colon\N\to\Q_2$ \emph{вполне округлена}, если
$\mu(x)$ является кратчайшей двоичной дробью в интервале $(\mu(x-1)$,
$\mu(x+1))$, а также $-\log\mu(\{x\})=O(\|x\|)$.
 \end{definition}

Последнее условие добавлено для удобства; его можно обеспечить, смешав
(монотонную) функцию $\mu$ с каким-нибудь простым распределением.

\begin{lemma*}
Всякая вычислимая функция $\mu\colon\N\to\Q_2$ может быть эффективно
преобразована во вполне округленную функцию $\mu_1$, вычислимую с
замедлением $($по сравнению с $\mu(x)${}$)$ в $\|x\|$ раз. При этом,
если $\mu$ монотонна $(${т.е.} $\mu'\ge 0${}$)$, то $\mu'_1\ge\mu'/4$.
\end{lemma*}

Монотонность обеспечивается сравнением $\mu(x)$ с $\mu(y)$ для всех $y$,
являющихся началами $x\in \Q_2$. Далее утверждение леммы достигается за
счет округления. Сначала мы округляем $\mu(x)$ до кратчайшей двоичной
дроби $p$, которая ближе к $\mu(x)$, чем к любому другому $\mu(y)$; эти
округленные значения мы называем \emph{точками}. Затем находим все
\emph{щели}, {т.е.} ближайшие к $p$ двоичные дроби всех меньших двоичных
длин. Затем для каждой щели (в порядке возрастания длины) находим точку,
которая ее заполняет при последовательных округлениях; так делается до
тех пор, пока щель для $x$ не будет найдена.

Вполне округленные функции $\mu$ обладают любопытным свойством: оба
числа $m(x)=\mu(x)/\mu(\{x\})$ и $-\log\mu(\{x\})={\|m(x)\|}$ всегда
целые, и потому $\mu(x)$ есть конечная двоичная дробь, у которой целая
часть нулевая, а после запятой идет $m(x)$. Поэтому $m$ распределено
почти равномерно: $2k\mu(m^{-1}(k))\in[1,2]$ при $k\in m(\N)$. Кроме
того, оно вычислимо за полиномиальное время, так же как и $m^{-1}$
(двоичный поиск). Следовательно, $m(x)$ можно рассматривать как
альтернативное представление для $x$, в котором распределение $\mu$
становится достаточно равномерным.

Вообще говоря, не всегда можно ограничиться простыми распределениями на
входах. Возможно, исходные данные $r$, использованные при построении
входа $x$, и имели простое распределение, но сам процесс $A$
преобразования $r$ в $x$ мог быть чем-то вроде односторонней функции. Мы
можем предполагать, что $A$ есть алгоритм с не слишком большим временем
работы, но не что распределение вероятностей на его выходах просто.
Возникающее на выходе $A$ распределение называется \emph{реализуемым}
(samplable). В \cite{18} такие распределения сводятся к равномерным, так
же как и рассматриваемые в данном пункте, хотя и с помощью другого
трюка.

\section{Полнота}

\textbf{4.1. Полные распределения и инверторы.} Что значит, что данная
функция трудна для обращения? Это можно уточнить двояко. Можно считать
функцию трудной, если трудные для обращения значения порождаются с не
слишком малой вероятностью. А можно требовать большего: чтобы
вероятность получения легкого для обращения значения была пренебрежимо
мала. Существуют различные способы свести одну задачу к другой, и мы
будем рассматривать первую из упомянутых задач.

Прежде всего отметим, что лемма позволяет перечислить все вычислимые за
время $t(x)$ распределения, сохраняя $t$ с точностью до линейного
множителя. Сложив все такие распределения с коэффициентами, образующими
сходящийся ряд (например, $1/i^2$), мы получим распределение в классе
$\hbox{TIME}(t(x)\|x\|)$, являющееся полным для класса
$\hbox{TIME}(t(x))$. Можно было бы соединить распределения всех
сложностей в одно, при этом каждое значение порождается с тем меньшей
вероятностью, чем больше сложность его порождения (как это делалось в
п.~3.1). Но мы предпочитаем иметь дело прямо с реализуемыми
распределениями.

 \begin {definition} Распределения вероятностей на выходе {\rm L}-ал\-го\-рит\-мов
без входа называются реализуемыми. Аналогичным образом {\rm L}-алгоритм
со входом задает реализуемое семейство распределений $($вход является
параметром$)$. \end{definition}

Обычно дается менее ограничительное определение, разрешающее больший
класс алгоритмов (более близкий к $\mathrm{LV}$ с полиномиальным
ограничением) и полиномиально б\'ольшие вероятности; мы рассматриваем
лишь L-алгоритмы, стремясь к большей точности.

\begin{proposition}\label{opt} Существует полное $($наибольшее с точностью до
постоянного множителя$)$ реализуемое семейство распределений.
\end{proposition}

В самом деле, L-алгоритмы можно перечислять, и полное реализуемое
семейство распределений можно получить, выбирая случайный L-алгоритм и
выполняя его. Описанный алгоритм требует в среднем времени $O(1)$ и
имеет не меньше шансов (с точностью до постоянного множителя) породить
``сюрприз'', чем любой другой\linebreak L-алгоритм. (У LV-алгоритмов с
полиномиальным ограничением вероятность ``не-\linebreak приятного
сюрприза''\ может быть больше, чем у описанного алгоритма, но различие
не более чем полиномиально.)

Полное распределение определено лишь с точностью до ограниченного
множителя, поэтому только в логарифмической целочисленной шкале оно дает
объективную меру трудности попадания в множество $X$ при данном значении
параметра $x$, определенную с точностью до ограниченного числа делений
шкалы.

О~б~о~з~н~а~ч~е~н~и~е.~ Через $\mathrm{Kl}(X/x)$ мы обозначаем $-\log_2
p(X/x)$, где $p(X/x)$ есть вероятность попадания в множество $X$
относительно полного семейства реализуемых распределений с параметром
$x$.

Как часто бывает, средство для атаки помогает найти и защиту.
Оптимальные алгоритмы поиска (указанные в \cite{19,20,21}) при нашем
определении получаются сами собой: полный генератор трудных задач
превращается в оптимальный алгоритм их решения. Напомним, что мы
``конвертируем''\ время работы в вероятность успеха, переходя к
L-алгоритмам, и измеряем их производительность этой вероятностью.
Алгоритм, порождающий наибольшее реализуемое распределение (с параметром
$x$) имеет наибольшую (с точностью до постоянного множителя) вероятность
$1/S(f/x)=2^{-\mathrm{Kl}(f^{-1}(x)/x)}$ порождения решений. Величина
$s(f/x)=\mathrm{Kl}(f^{-1}(x)/x)$ характеризует трудность конкретного
примера $x$ и может быть названа его \emph{стойкостью}. Наш оптимальный
генератор в среднем требует $O(1)$ шагов на один запуск и $S(f/x)$
запусков. Никакой другой алгоритм не может дать лучшего результата.

О~т~к~р~ы~т~а~я~ п~р~о~б~л~е~м~а.~ Постоянный множитель в оптимальном
алгоритме обращения может быть произвольно большим. Неизвестно, можно ли
ограничить этот множитель (для достаточно длинных входов) некоторой
абсолютной константой, не зависящей от выбора сравниваемого с
оптимальным алгоритма (скажем, числом $10$).

\textbf{4.2. Задачи обращения и односторонние функции.} Полное
распределение дает не меньшую вероятность получить трудный вход, чем
любое другое. Благодаря этому в качестве трудной для обращения функции
можно взять любую NP-полную функцию: все они одинаково хороши. Однако
обычно хочется найти функцию, которую трудно обратить для какого-либо
обычного (например, равномерного) распределения на входах. Неожиданным
образом лемма как раз и указывает кодирование, преобразующее заданное
распределение (вычислимое за полиномиальное время) в равномерное. С его
помощью любая NP-полная функция становится максимально трудной для
обращения. Однако в сочетании с таким кодированием функция теряет
привлекательность, так что вопрос о построении привлекательной функции,
трудной для обращения при обычном распределении на входах, сохраняется.

Как известно, ни для одной функции не удалось доказать, что она трудна
для обращения, хотя многие функции кажутся таковыми. В
\cite{17,18,22,23,24,25} (и ряде других работ) указан ряд комбинаторных
и алгебраических задач, которые являются в среднем полными при
равномерном распределении входов ({т.е.} они не проще любой задачи
обращения с реализуемым распределением).

Однако эти результаты все еще не дают односторонних функций. Разница
между односторонними функциями и трудными в среднем задачами обращения
может быть выражена многими способами. Простейший из них состоит в том,
чтобы определить одностороннюю функцию как трудную в среднем задачу
обращения функции, \emph{сохраняющей длину}. В этом случае различие
между выбором случайного аргумента или случайного значения (существенное
для односторонних функций) перестает быть важным.

В самом деле, каждому решению соответствует только одно значение,
поэтому для сохраняющих длину функций вероятность появления значения
функции из данного множества значений, имеющих решения, не меньше
вероятности, соответствующей равномерному распределению. При этом,
однако, может быть много ``близнецов''~-- решений, соответствующих
одному и тому же значению. В этом случае мы можем модифицировать функцию
следующим образом. Отгадаем логарифм числа близнецов для данного решения
$w$ (пусть этот логарифм равен $k$) и рассмотрим случайный элемент $a$
универсального семейства хеш-функций $h_a(w)$. Будем считать выходом
функции набор $f(w),k,a,h'_a(w)$, где $h'$ получается из $h$, если
оставить только $k$ первых битов. (Эти значения на $k$ битов длиннее
решений и отображаются с помощью другой хеш-функции в строки той же
длины, что и решения.) Содержащаяся в этом выходе дополнительная
информация (при правильно угаданном значении $k$) близка к случайной, и
потому не помогает обращению. С другой стороны, близнецы разделяются на
небольшие группы, и потому количество (и вероятность при равномерном
распределении) трудных значений и количество их прообразов становятся
сравнимыми. Обратное утверждение также верно:

\begin{proposition}
Любая односторонняя функция с мультимедианой $V(k)$ времени оптимального
обращения $(${т.е.} $S(x)${\rm,} для $x=f(w)$ и равномерно
распределенного $w${}$)$ может быть преобразована в сохраняющую длину
одностороннюю функцию, у которой для $1/O(k)$ доли примеров надежность
полиномиально связана с $V$. \end{proposition}

Прежде всего, мы увеличиваем долю трудных для обращения значений, как
описано в конце п.~3.2. Если число трудных значений существенно меньше,
чем число их прообразов, функцию все равно можно переделать в
сохраняющую длину без изменения трудности, разделяя близнецов, как
описано в предыдущем абзаце. Более подробно процесс применения
хеширования к односторонним функциям исследован в~\cite{18}.

\textbf{4.3. Полная односторонняя функция: продолжение замощения.} Мы
сейчас покажем, как модифицировать задачу о замощении, получив из нее
комбинаторную полную одностороннюю функцию. Такие функции не встречались
в литературе, хотя в~\cite{9} приводится построение искусственной полной
односторонней функции. Конечно, хорошо бы найти несколько примеров
полных односторонних функций, менее искусственных, {т.е.} выглядящих
привлекательно для человека, не знакомого (и не желающего знакомиться) с
теорией алгоритмов. Мы сейчас приведем один такой пример ``для
затравки'', надеясь на то, что рано или поздно будет накоплена
``критическая масса''\ полных функций, свед\'ение которых позволит
доказывать полноту разнообразных интересных кандидатов в односторонние
функции.

\textbf{NP-полнота и OWF-полнота.} Широкий успех доказательств
NP-полноты многочисленных комбинаторных задач остается загадочным. Это
вопрос скорее искусства, чем науки, и потому не требующий однозначного
объяснения. Но одной из причин, видимо, является большой набор готовых
NP-полных комбинаторных задач, описание которых не требует анализа
(утомительных) деталей, характерных для \emph{детерминированных}
вычислительных моделей. Для \emph{полноты в среднем} таких примеров пока
существенно меньше, хотя они и накапливаются. С другой стороны, мой
вопрос о построении явного простого примера полной односторонней функции
оставался безответным в течение двух десятилетий.

Полная односторонняя функция может быть получена модификацией
универсальной машины Тьюринга (UTM). В свою очередь, протоколы работы
UTM могут быть легко преобразованы в комбинаторные объекты с
определенными свойствами (типа замощений). Описание этих свойств проще
описания машины Тьюринга, поскольку теперь нет необходимости заботиться
о детерминированности вычисления: отношение, которое мы строим, и не
должно быть детерминированным. Упрощенные аналоги вычислений
привлекательны своей простой комбинаторной структурой, которая позволяет
свести возникающую задачу ко множеству других (тем самым доказав полноту
последних).

Этот подход действительно позволяет построить NP-задачи, являющиеся
полными в среднем (average-complete). Однако он не позволяет обеспечить
условие сохранения длины, которое существенно при построении полных
односторонних функций. (Оно может быть заменено другими требованиями, но
и эти требования не удается удовлетворить в описанной выше конструкции.)
Сейчас мы покажем, как можно сравнительно простыми средствами (понятие
расширения) попытаться преодолеть возникающие проблемы. При этом мы
обеспечим сохранение длины и простую комбинаторную структуру замощения.
Мы надеемся, что эта полная односторонняя функция может быть полезна как
исходная точка для доказательств полноты интересных односторонних
функций с помощью сведений.

\def\tilet {Плитки: единичные квадраты, углы которых помечены буквами;
 их можно прикладывать сторона к стороне, если буквы совпадают.}
 \def\expan {Расширение: максимальное продолжение заданного частичного
замощения квадрата с отмеченной границей в таком порядке, при котором
каждая следующая плитка определяется {\bf однозначно} (при заданном
наборе плиток).}

\newlength\tilel\setlength\tilel{\columnwidth}
\newlength\tbx\setlength\tbx{10ex}
\addtolength\tilel{-\tbx}\addtolength\tilel{-1em}
\medskip\noindent\fbox{\parbox{\tilel}{\tilet\ \expan}
 \hfill\parbox{\tbx} {\mbox
 {\fbox{\parbox{3ex} {\makebox[3ex]{a\hfill x} \makebox[3ex]{e\hfill
	 r}}}
  \fbox{\parbox{3ex} {\makebox[3ex]{x\hfill c} \makebox[3ex]{r\hfill
	z}}}}
  \par\vspace {.5ex} \mbox
 {\fbox{\parbox{3ex} {\makebox[3ex]{e\hfill r} \makebox[3ex]{n\hfill
	 s}}}
  \fbox{\parbox{3ex} {\makebox[3ex]{r\hfill z} \makebox[3ex]{s\hfill
	z}}}}}}

\begin{definition}
Расширением замощения мы называем следующую функ-\linebreak цию:
аргумент~-- верхняя строка замощения и набор разрешенных плиток;
значение~-- нижняя строка замощения, получаемого расширением верхней
строки, и набор разрешенных плиток.\end{definition}

\begin{theorem*}
Функция расширения замощения является односторонней тогда и только
тогда, когда односторонние функции существуют. \end{theorem*}

\emph{Сведение:} Мы начинаем с универсальной машины Тьюринга и добавляем
к ней счетчик, который прерывает ее работу после, скажем, $n^2$ шагов.
Мы сохраняем копию программы (начальный отрезок входа) неизменной, а
также принудительно делаем длину выхода равной длине входа. Эта
конструкция дает нам полную одностороннюю функцию, сохраняющую длину
(правда, описываемую с помощью вычислительной модели). Далее мы сводим
вычисление UTM (с указанными модификациями) к задаче о замощении с
помощью стандартного приема. Мы добавляем специальный \emph{граничный
символ} и разрешаем его лишь в плитках, в которых он сочетается с
символами входного или выходного алфавитов (одинакового размера), а
также с \emph{символом конца ленты} или с символом, начинающим
вычисление (в зависимости от стороны плитки). Остается воспользоваться
определением расширения.

Замощение является простой комбинаторной задачей, но ее
недетерминированная природа вынуждает нас указывать все плитки в
квадрате, что мешает сохранению длины. Если требовать, чтобы набор
плиток вынуждал детерминизм вычисления, получится громоздкая
конструкция, которую трудно связать с простыми комбинаторными задачами.
Вместо этого, говоря о расширениях, мы не накладываем ограничений на
множество разрешенных плиток, зато разрешаем прикладывать лишь плитки,
которые однозначно определяются (при данном наборе разрешенных плиток)
уже имеющимися. При этом некоторые частичные замощения квадрата удается
продолжить до полных, добавляя плитки одну за другой, другие~-- нет.
Этот процесс приводит к потере эффективности (небольшой для параллельных
моделей), но это для нас не важно.

Остается интересная задача: свести эту одностороннюю функцию к другим
простым комбинаторным или алгебраическим функциям, тем самым доказав
их\linebreak полноту.

Произвольные односторонние функции не так просто применить на практике.
Во многих случаях (например, при построении псевдослучайных
последовательностей) других предположений формально не требуется, но
приходится использовать построения, которые катастрофически ухудшают
количественные показатели эффективности. Более пригодные на практике
конструкции используют односторонние функции с некоторыми
дополнительными свойствами, например, с малой энтропией в смысле Реньи.
Это же требование используется при преобразовании слабо односторонней
функции в сильно одностороннюю (с сохранением параметра стойкости, как
описано в~\cite{26}). Следующее замечание указывает один из возможных
путей получения односторонней функции, удовлетворяющей этому
требованию.\linebreak (В нем выражение $f(x)+ax$ может быть заменено на
другие хеш-функции.)

\emph{Замечание}. Аргументы функции $g(a,x)=(a,f(x)+ax)$ в среднем имеют
не более одного близнеца для любой сохраняющей длину функции $f$ и для
$a,x\in\mathrm{GF}_{2^{\|x\|}}$.

Г~и~п~о~т~е~з~а.~ \emph{Построенная таким образом функция $g$ является
односторонней, если функция $f$ была таковой, и имеет тот же $($с
точностью до полиномиального множителя$)$ параметр стойкости.}